\documentclass[amsmath,amssymb,aps,prd,11pt,tightenlines,superscriptaddress,nofootinbib,preprintnumbers,notitlepage]{revtex4}

    \usepackage{multirow}
    \usepackage{graphicx}
    \usepackage{tabularx}
    \usepackage{mathtools}
    \usepackage{slashed}
    \usepackage{booktabs}
    \usepackage{url}
    \usepackage[maxfloats=100]{morefloats}
    \usepackage{color,xcolor}
    \usepackage[normalem]{ulem}
    \usepackage{subfig}
    \definecolor{dark-red}{rgb}{0.8,0.0,0.0}
    \definecolor{dark-blue}{rgb}{0.0,0.0,0.8}
    \definecolor{dark-green}{rgb}{0.,0.6,0.}
    \usepackage{todonotes} 
    \usepackage{xspace} 

    \makeatletter
    \def\l@subsubsection#1#2{}
    \makeatother
    
    \usepackage{enumitem}
    \setlist[itemize]{itemsep=1pt,parsep=1pt, topsep=1pt}
    \newcommand{\VLL}{VLL\xspace}
    \newcommand{\VLLL}{\ensuremath{\mathrm{L}}}
    \newcommand{\VLLE}{\ensuremath{\mathrm{E}}}
    \newcommand{\VLLN}{\ensuremath{\mathrm{N}}}
    \newcommand{\PB}{\ensuremath{\mathrm{B}}}
    \newcommand{\MGMCatNLO}{MadGraph5\_aMC@NLO} 
    \newcommand{\pythia}{{\sc Pythia}}
    \newcommand{\delphes}{{\sc Delphes}}
    \newcommand{\fastjet}{{\sc FastJet}}
    \newcommand{\met}{\ensuremath{\slashed{E}}}
    \newcommand{\fbinv}{\mbox{\ensuremath{~\mathrm{fb^{-1}}}}}
    \newcommand{\abinv}{\mbox{\ensuremath{~\mathrm{ab^{-1}}}}}
    \newcommand{\PZ}{\ensuremath{\mathrm{Z}}}
    
    \newcommand{\PW}{\ensuremath{\mathrm{W}}}

    \newcommand{\pt}{\ensuremath{p_\mathrm{T}}}

    \newcommand{\al}{{\it et al. }}
    \usepackage{ragged2e}

\begin{document}
    
    
    \title{Vector-like Lepton Searches at a Muon Collider in the Context of the 4321 Model}

    \author{Qilong Guo}
    \affiliation{School of Physics and State Key Laboratory of Nuclear Physics and Technology, Peking University, Beijing, 100871, China}
    \author{Leyun Gao}
    \affiliation{School of Physics and State Key Laboratory of Nuclear Physics and Technology, Peking University, Beijing, 100871, China}
    \author{Yajun Mao}
    \affiliation{School of Physics and State Key Laboratory of Nuclear Physics and Technology, Peking University, Beijing, 100871, China}
    \author{Qiang Li}
    \affiliation{School of Physics and State Key Laboratory of Nuclear Physics and Technology, Peking University, Beijing, 100871, China}
    
    \date{\today}
    
    \begin{abstract}
A feasibility study is performed for searching vector-like leptons ({\VLL}s) at a muon collider, in the context of the ``4321 model'', an ultraviolet-complete model with rich collider phenomenology together with potential to explain recent existing some {\PB} physics measurements or anomalies. 
Pair productions and decays of {\VLL}s lead to interesting final state topology with multi-jets and multi-tau leptons. 
In this paper, we perform a Monte Carlo study with various machine learning techniques, and examine the projected sensitivity on vector-like leptons over a wide mass range at a TeV-scale muon collider. 
We find that a 3 TeV muon collider with only $10\ \mathrm{fb}^{-1}$ of data can already be sensitive to cover the mass range of a vector-like lepton up to 1450 GeV in the ``4321 model''.
    \end{abstract}
    
    \maketitle
    \tableofcontents
    
    \section{Introduction}
    \label{sec:intro}
The 4321 model~\cite{DiLuzio:2017vat, DiLuzio:2018zxy, Greljo:2018tuh, Calibbi:2017qbu, Blanke:2018sro} is an ultraviolet-complete model that extends the standard model (SM) gauge groups to a larger $\text{SU}(4) \times \text{SU}(3)' \times \text{SU}(2)_L \times \text{U}(1)'$ group. It is motivated by recent measurements of B hadron decays that are in tension~\cite{DiLuzio:2017vat} with the SM.
It can provide a combined explanation for multiple anomalies observed in b hadron decays, which point to lepton flavour nonuniversality.
This model also provides a novel interesting final state topology to search at colliders, with multi-jets and multi-tau leptons. 
Searches for pair production
of the lightest new particles in this model, the vector-like leptons ({\VLL}s) has recently been performed~\cite{CMS:2022cpe} at the CMS experiment based on the data collected in 2017 and 2018 corresponding to a total integrated
luminosity of 96.5\fbinv. Interestingly, a mild excess has been found at the level of 2.8 standard deviations, for a representative VLL mass point of 600 GeV~\cite{CMS:2022cpe}. In this paper, we are interested in a similar study at the future TeV-scale Muon collider~\cite{Daniel20}, which should have great potential to explore {\VLL}s and other predicted new particles in the 4321 model to a large extent.

A muon--muon collider with the center-of-mass energy at the multi-TeV scale has received much-revived interest~\cite{Daniel20} recently, which has several advantages compared with both hadron--hadron and electron--electron colliders~\cite{Mario16, Antonio20, Dario18}. As massive muons emit much less synchrotron radiation than electron beams, muons can be accelerated in a circular collider to higher energies with a much smaller circumference. On the other hand, because the proton is a composite particle, muon--muon collisions are cleaner than proton--proton collisions and thus can lead to higher effective center-of-mass energies. However, due to the short lifetime of the muon, the beam-induced background (BIB) from muon decays needs to be examined and reduced properly. Based on a realistic simulation at $\sqrt{s}=1.5$~TeV with BIB included, Ref.~\cite{Nazar20} found that the coupling between the Higgs boson and the b-quark can be measured at percent level with order\abinv\ of collected data.
    
\section{Physics processes}
\label{sec:muc}

\begin{figure}[ht]
    \centering
    \includegraphics[width=0.6\textwidth]{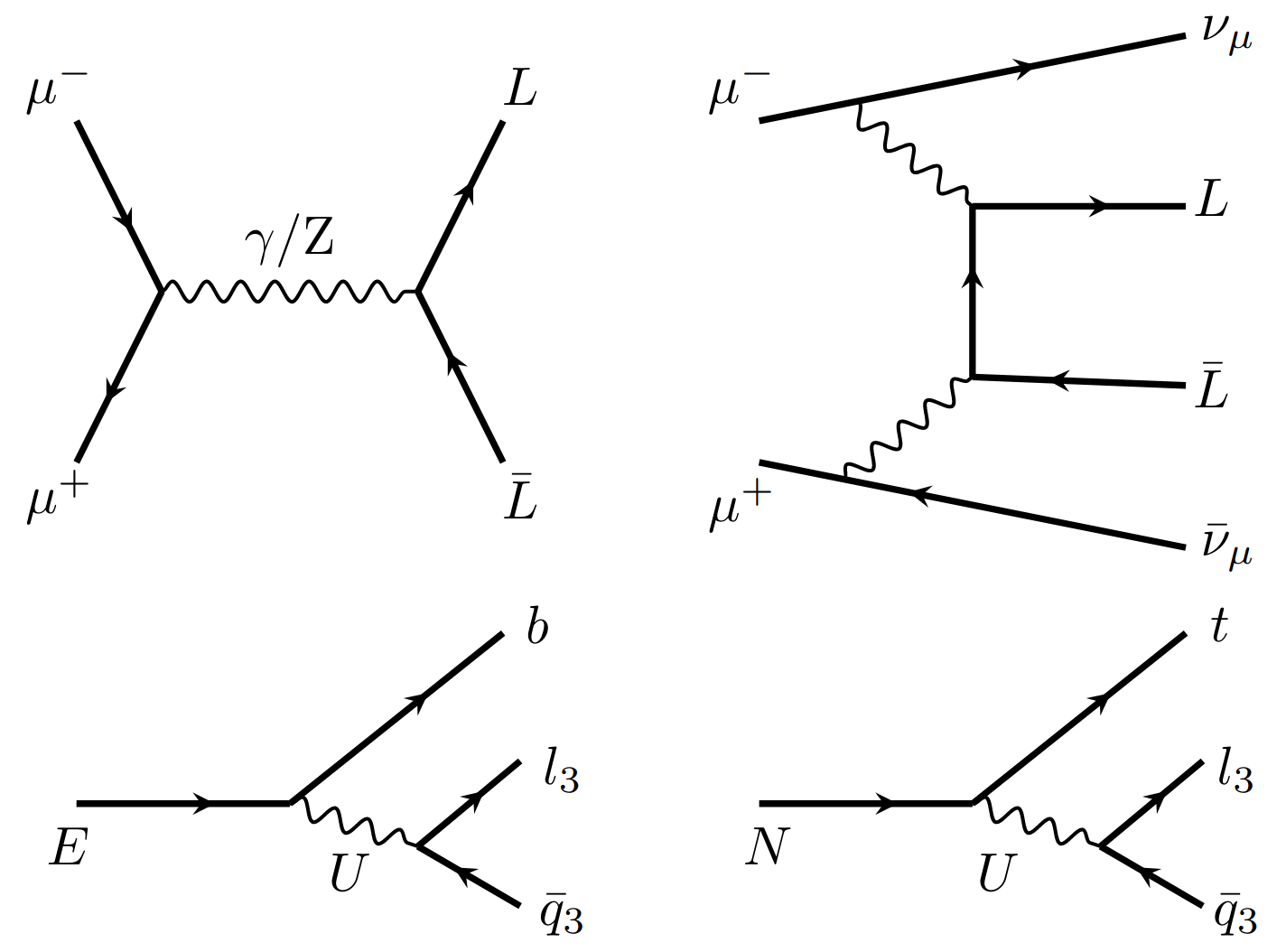}
    \caption{Top side: example Feynman diagrams showing \VLL 
     pair productions through $s$-channel and vector boson scattering at a TeV muon collider. {\VLLL} represents either the neutral \VLL, {\VLLN}, or the charged \VLL, {\VLLE}. Bottom side: vector-like lepton decays are mediated by a vector leptoquark, $U$. These decays are primarily to third-generation leptons and quarks.}
    \label{fig:feynman_diagrams}
\end{figure}

In the 4321 model, {\VLL}s come in electroweak doublets with one charged \VLL, {\VLLE}, and one neutral \VLL, {\VLLN}, whose masses are taken to be equal. The {\VLL}s may be produced via electroweak production through their couplings to the SM {\PW} and {\PZ}/$\gamma$ bosons or via interactions with a new heavy $\rm{Z}^\prime$ boson introduced by the 4321 model. In this paper, we consider only the electroweak production and ignore potential contributions from the $\rm{Z}^\prime$ boson.  Examples of Feynman diagrams showing \VLL pair productions through a) $s$-channel ($\mu^+\mu^-\rightarrow \VLLL\VLLL$ with {\VLLL} represents either {\VLLN} or {\VLLE}) and b) vector boson scattering (vbs) at a TeV muon collider ($\mu^+\mu^-\rightarrow \VLLL\VLLL \nu_\mu\bar{\nu}_\mu$), as well as diagrams of the {\VLL} decays, are shown in Fig.~\ref{fig:feynman_diagrams}. 

    \begin{figure}[ht]
    \includegraphics[width=.6\textwidth]{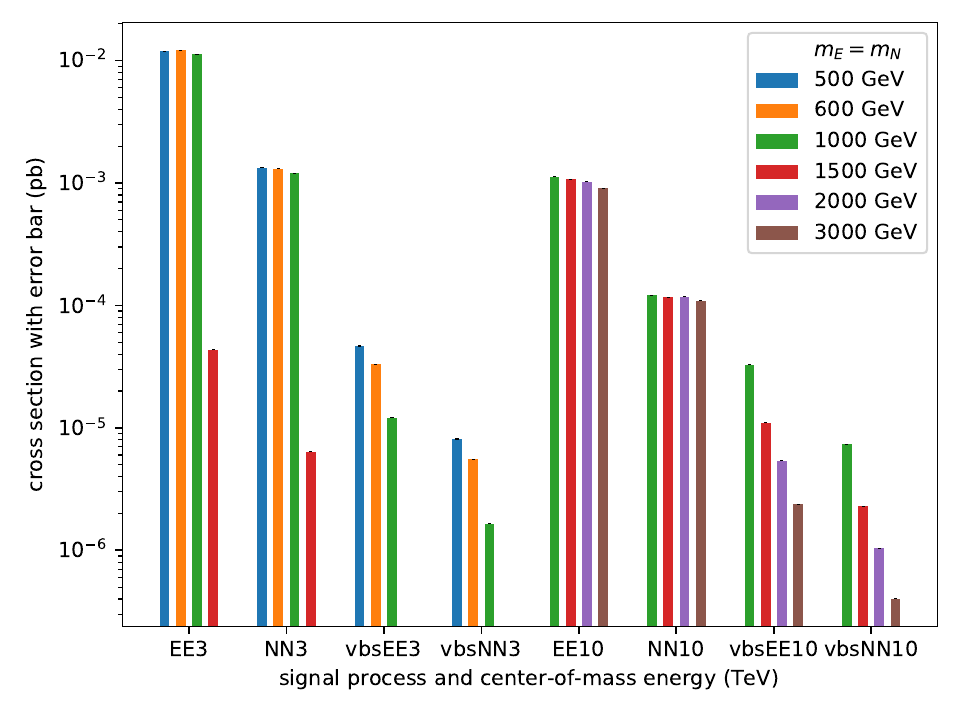}
    \caption{Cross sections of various signal processes and \VLL~mass points in two benchmark collider proposals with  $\sqrt{s}=3$ or 10 TeV. The x-axis shows the process symbol (pair productions, $\mu^+\mu^-\rightarrow \VLLL\VLLL$, or, vbs productions, $\mu^+\mu^-\rightarrow \VLLL\VLLL \nu_\mu\bar{\nu}_\mu$) and the $\sqrt{s}$ in the scale of $\mathrm{TeV}$, while the y-axis on the logarithmic scale indicates the cross sections in the unit of $\mathrm{pb}$.\;The VBS processes with $m_E = m_N = 1500\ \mathrm{GeV}$ and $\sqrt{s}=3$ TeV result in negligible cross sections.
    }
    \label{fig:cs-all}
    \end{figure}

Fig.~\ref{fig:cs-all} illustrates cross sections of various processes and \VLL~mass points at center-of-mass (C.O.M) energy $\sqrt{s}=3$ or 10 TeV. The cross sections of EE processes are larger than the second ones, NN processes, by one order of magnitude. The VBS cross sections are smaller by several orders of magnitude than the pair productions. Hence we will then focus on the EE pair production processes afterward.

The background processes we consider include:
 \begin{itemize}
\item b1) $\mu^+\mu^-\rightarrow b\bar{b}\,, b\bar{b}Z$,
\item b2) $\mu^+\mu^-\rightarrow \tau^+\tau^-\,, \tau^+\tau^-Z$,
\item b3) $\mu^+\mu^-\rightarrow W^+W^-\,, W^+W^-Z$,
\item b4) $\mu^+\mu^-\rightarrow t\bar{t}\,, t\bar{t}Z$,
\item b5) vbs productions of $b\bar{b}\,,\tau^+\tau^-\,,W^+W^-\,,t\bar{t}$.
 \end{itemize}

\section{Simulation and analysis framework}
\label{sec:simulation}



We consider the muon collider benchmarks in which $\sqrt{s}=3$ TeV and L = ab$^{-1}$ in this study.
Both signal and background events are simulated with \MGMCatNLO, then showered and hadronized by \pythia8~\cite{Sjostrand:2014zea}. 
The final state jets are clustered using \fastjet~\cite{Cacciari:2011ma} with the VLC~\cite{BORONAT201595,Boronat:2016tgd} algorithm at a fixed cone size of $R_{\rm jet}=0.5$. 
We use \delphes~\cite{deFavereau:2013fsa} 3.5.0 to simulate detector effects with the newly-added card for the muon collider detector~\cite{mucard}. 
%
The jet flavor tagging efficiencies are listed in Table~\ref{tab:jet_eff}. 
Notice however currently the jet tagging techniques for muon colliders are in a preliminary stage~\cite{muon_btag_1, Nazar20} and have a large potential to improve.

\begin{table}[htbp]
    \centering
    \begin{tabular}{|c|c|c|}
    \hline
        Flavour & Tagged as light & Tagged as $b$ \\
        \hline
        light & 0.9 & 0.01 \\
        \hline
        $b$ & 0.1 & 0.7 \\
        \hline
    \end{tabular}
    \caption{Flavour tagging efficiencies used in this analysis.}
    \label{tab:jet_eff}
\end{table}

The hadronic $\tau$ identification efficiency is chosen from the performance of the CMS detector. The mis-tagging rates are both $\sim1\%$ from CEPC detector simulation~\cite{cepc_reco} and CMS estimation~\cite{tau_reco}.

\begin{figure}[htb]
    \centering
    \includegraphics[width=0.32\textwidth]{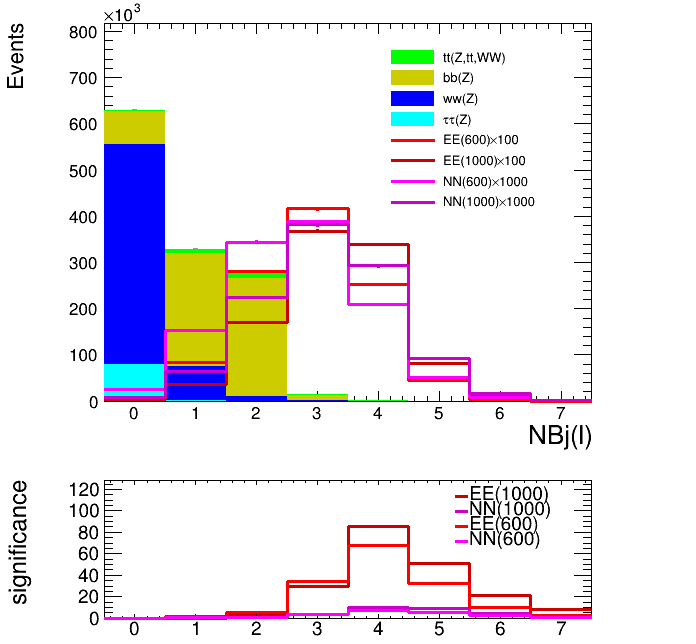}
    \includegraphics[width=0.32\textwidth]{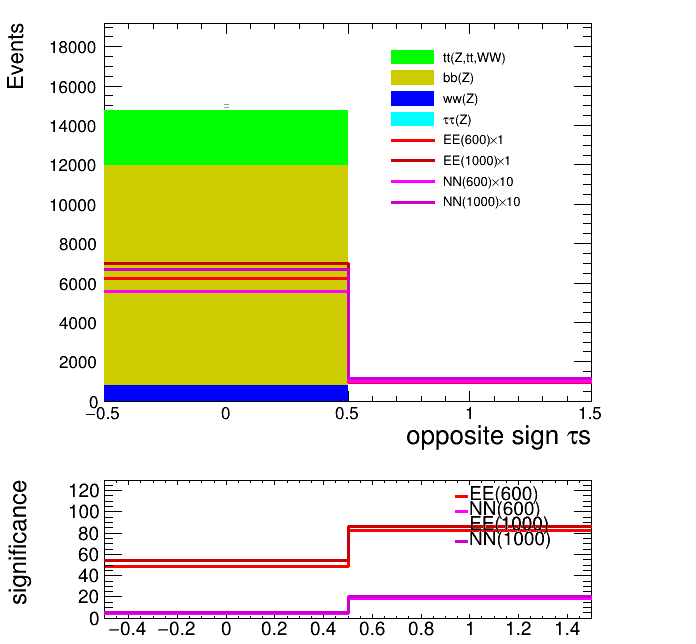}
    \includegraphics[width=0.32\textwidth]{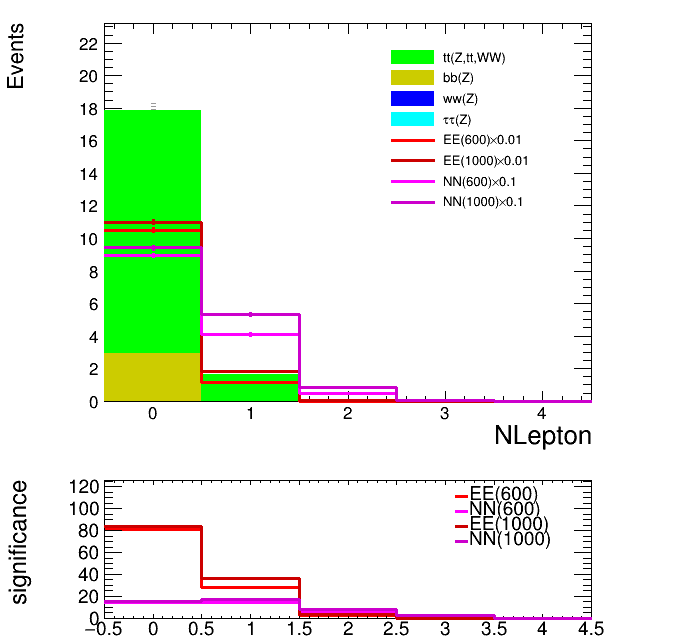}
    \caption{Signal and stacked background distributions on multiplicities of identified b-jet , tau (requiring 3 b-tagged jets in events) and additional leptons (requiring 3 b-tagged jets in events and no additional leptons in the events), simulated at a 3 TeV muon muon collider. The ratio plots below show the statistic only significance estimated for each signal points, based on which one can decide the selections with the best performance. }
    \label{fig:pre}
\end{figure}

Fig.~\ref{fig:pre} shows signal and stacked background distributions on multiplicities of identified tau,  b-jet and additional leptons, simulated at a 3 TeV muon collider. One can see pair productions of EE processes have the dominate significance. Based on the optimization performance as shown in Fig.~\ref{fig:pre}, we define our preselection criteria as below: events must include at least a pair of identified opposite sign $\tau$s, 3 b-tagged jets, while events with additional leptons are vetoed.

\section{Analysis Methods and results}
\label{sec:methods}

%
%
%
The reconstruction of the \VLL candidates starts with selecting and clustering five or six jets into two reconstructed {\VLL}s using the following algorithm looping through at most 10 jets:
\begin{itemize}
    \vspace{0.15cm}
    \item Select the 2 $\tau$ jets and 4 btagged jets ( For events with only 3 btagged jets, a jet is selected or zero padding is applied if there is no jet expect the jets already considered) as b candidates. 
    \item Construct the combinatorics of 2 $\tau$ and 4 b candidates into (${\tau}_1,b_1,b_2$),(${\tau}_2,b_3,b_4$).
    \vspace{0.15cm}
    \item Calculate $\Delta M^2=M({\tau}_1,b_1,b_2)^2-M({\tau}_2,b_3,b_4)^2$,
    \vspace{0.15cm}
    \item (${\tau}_1,b_1,b_2$) and (${\tau}_2,b_3,b_4$) giving lowest $\Delta M^2$ are reconstructed as 2 VLLs.
    \vspace{0.15cm}
\end{itemize}
Addtional pairs of VLLs are reconstructed by requiring the relatively small $\Delta M^2$. This algorithm is used to construct the input for BDT training.

\subsection{BDT}
For the implementation of the BDT, we shuffle the signal and background events and define the training and test sets with the event ratio of $4:1$. The BDT with 850 trees and a maximum depth of 3 is trained. We apply the per-event weight during the training to account for the cross-section difference among the background processes. The weight is defined by
\begin{equation}
    n_{L_X}=\sigma_XL/N_{G_X},
\end{equation}
where $\sigma_X$ denotes the cross-section of a process, $L$ denotes the default target luminosity in this study as around 1\abinv\ for a 3 TeV muon collider, and $N_{G_X}$ denotes the generated number of events. The total signal yields are reweighted to match that of the total background during training for the robustness of the trained model. Signal yields corresponding to different VLL mass are weighted to be the same. 
The input features, namely reconstructed kinematics of each event used in training, are listed as follows and summarized in TABLE~\ref{tab:BDT_input}. 

\begin{itemize}
\vspace{0.15cm}
\vspace{0.15cm}
\item $(\pt,\,\eta,\,\phi,\,mass)$ of at most 10 jets including $\tau$s. $\tau$ jets are prior to $b$ candidates and both of them has higher priorities then rest jets.
\vspace{0.15cm}
\item $(\pt,\,\eta,\,\phi)$ of the missing energy \met,
\vspace{0.15cm}
\item $(\pt,\,\eta,\,\phi,mass,Nbjets\,in\,VLL)$ of the 6 pair of VLL canidates,
\vspace{0.15cm}
\item $(\Delta \phi,\,\Delta R)$ between each pair of VLLs candidates 
\end{itemize}

\begin{table}[htbp]
    \centering
    \caption[]{Summary of features used for the BDT training.}
    \label{tab:BDT_input}
    \begin{minipage}{0.8\textwidth}
    \begin{tabularx}{\textwidth}{X<{\centering}X<{\centering}p{2cm}<{\centering}}
     \hline\hline
     Objective&Features&Number of features\\
    \hline
        Each jets&$(\pt,\,\eta,\,\phi,mass,flavor)$&50\\
        \met&$(\pt,\,\eta,\,\phi)$&3\\
        Each pair of VLL candidates&$(\pt,\,\eta,\,\phi,mass,flavor,\Delta \phi,\,\Delta R)$&48\\
        Total:&&101\\
     \hline\hline
    \end{tabularx}
    \end{minipage}
\end{table}




\subsection{Simple Fully-connected DNNs}
Using variables of jets and MET from the input of the BDT model, for each signal or background event, we have a flattened feature vector of size 53 to describe the information extracted from the final state. A direct way to construct a classifier based on the already vectorized input without any concise symmetry is to build fully-connected dense layers (simple fully-connected DNNs).

As the input features spread over many orders of magnitude, feature-wise normalization is done before feeding the data into the network. Several architectures of simple fully-connected DNNs are tried, which share the same 53-neuron input layer and the same sigmoid-activated 1-neuron output layer, differ only with the number and the size of the ReLU-activated hidden layers. The RMSProp optimizer with the initial learning rate $10^{-4}$ and the batch size 32 is used to minimize the binary cross entropy loss. The training, validation, and test samples are randomly taken from the dataset following the $8:1:1$ ratio, and the training samples are shuffled at the beginning of each epoch. Validation is performed at the end of each epoch.

\subsection{The ABCNet Model}

The ABCNet~\cite{ABCNet} is an attention-based graph neural network, taking advantage of the graph attention polling (GAP)~\cite{GAPNet} layers to enhance the local feature extraction, in addition to the previous state-of-the-art model ParticleNet~\cite{ParticleNet}, for the tasks of, e.g., quark-gluon tagging and pileup mitigation. For the purpose of binary classification between the signal and background events in the context of the 4321 model in literature~\cite{CMS:2022cpe} as well as in this paper, the model takes as input batches of the point-cloud encoded events and outputs the predicted probabilities for each event to be signal, which we will call ABCNet scores in the following.

\begin{table}[htbp]
    \centering
    \caption{Input local features of our ABCNet model.}
    \begin{tabular}{rrl}
        \hline
        \hline
        \qquad No. & \qquad Feature & \qquad Description \qquad \\
        \hline
        \qquad 0 & \qquad $\eta$ & \qquad pseudorapidity of the jet momentum \qquad \\
        \qquad 1 & \qquad $\phi$ & \qquad azimuthal angle of the jet momentum \qquad \\
        \qquad 2 & \qquad $\log\frac{p_\mathrm T}{\mathrm{GeV}}$ & \qquad traverse momentum of the jet, with logarithmic scaling \qquad \\
        \qquad 3 & \qquad $\log\frac{m}{\mathrm{GeV}}$ & \qquad static mass of the jet, with logarithmic scaling \qquad \\
        \qquad 4 & \qquad $Q$ & \qquad charge for a $\tau$ jet or zero filling otherwise \qquad \\
        \qquad 5 & \qquad $\mathrm{BTag}_{50}$ & \qquad $b$-tagging result for a $b$ candidate with $\epsilon_\mathrm s = 50\%$ or zero filling otherwise \qquad \\
        \qquad 6 & \qquad $\mathrm{BTag}_{70}$ & \qquad $b$-tagging result for a $b$ candidate with $\epsilon_\mathrm s = 70\%$ or zero filling otherwise \qquad \\
        \qquad 7 & \qquad $\mathrm{BTag}_{90}$ & \qquad $b$-tagging result for a $b$ candidate with $\epsilon_\mathrm s = 90\%$ or zero filling otherwise \qquad \\
        \hline
        \hline
    \end{tabular}
    \label{tab:local-features}
\end{table}

Specifically, the point-cloud encoding contains both detailed information about jets (local features, including $\tau$ jets and $b$ candidates) and the one about the whole event (global features). For each event, the cloud collects at most 10 jets with their kinematics, charges, and $b$-tagging results. $\tau$ jets are prior to $b$ candidates and both of them are then sorted by $p_\mathrm T$ in descending order. Jets with lower priorities are discarded if more than 10 are found, and zero padding is applied to ensure 10 points to input otherwise. We summarise the local features in TABLE \ref{tab:local-features}, in addition to which we provide the missing energy as the global features. The input two categories of features are integrated by the ABCNet to predict the ABCNet score. We cut the events as follows before building the data set:
\begin{itemize}
    \item at least 3 $b$-tagged candidates with $\epsilon_\mathrm s = 90\%$;
    \item at least 2 $\tau$, with at least 1 $\tau^+$ and at least 1 $\tau^-$;
    \item no lepton.
\end{itemize}
Then the events are shuffled and split into the training, validation, and test set with the $8:1:1$ ratio.

\begin{figure}
    \centering
    \includegraphics[width=0.8\textwidth]{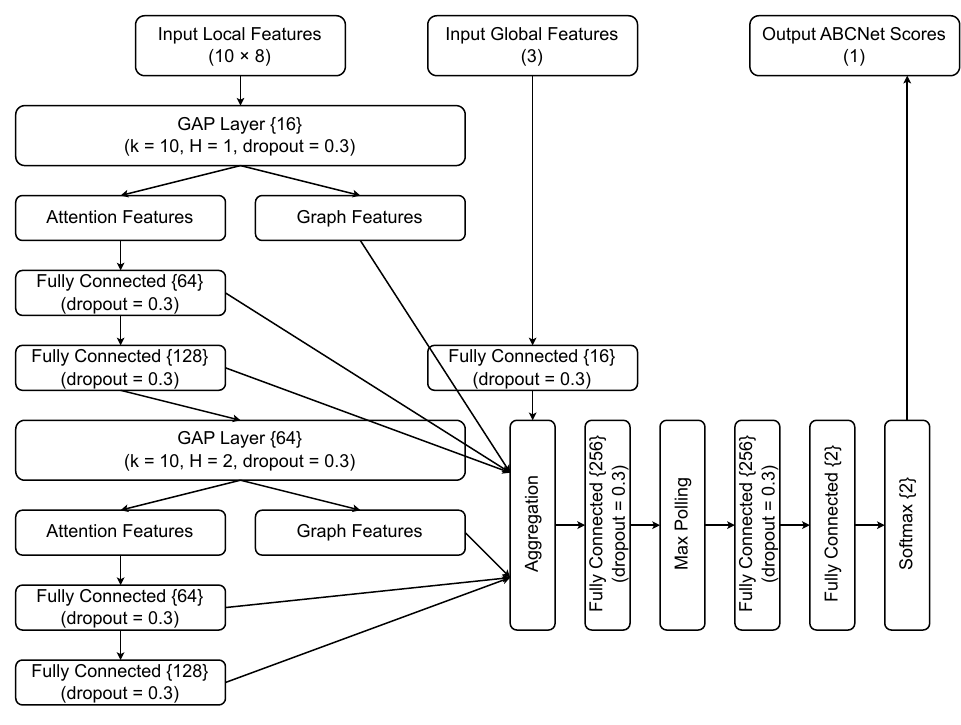}
    \caption{The ABCNet architecture used in this paper.}
    \label{fig:ABCNet}
\end{figure}

Our ABCNet model is similar to the one the CMS collaboration used in literature \cite{CMS:2022cpe}, except for the 3 input global features and the fully connected layer of size 16 connecting them to the aggregation. It is tuned starting from Fig. 1 in literature \cite{ABCNet}, shrinking the edge size of the first GAP block to 16, decreasing the capacities of the layers right after the two GAP blocks both to 64, and replacing the structures after the aggregation to two fully connected layers of size 256 sandwiching a max-polling layer before the softmax activated output layer of size 2. In addition, to avoid over-training, we apply a dropout after each hidden fully connected layer, whose ratio is tuned to be 0.3. FIG. \ref{fig:ABCNet} shows the architecture of the ABCNet we use in this paper. We use the Adam optimizer with batch size equaling 64 and the learning rate decaying exponentially from $10^{-2}$, which performs well both in the feature extraction and the over-fit avoidance. We monitor the training process with the metric validation accuracy and perform an early stop if the metric value doesn't improve over the nearest 10 epochs. The best-fit model with the highest validation accuracy is then tested over the test set to make accuracy and AUC evaluation.

\begin{table}
    \centering
    \caption{Performance of the ABCNet model over the 10 different folds and the overall values. In the table, std denotes the standard deviation, and S and B denote signal and background, respectively. The overall AUC is computed by gathering the 10 folds to evaluate the overall FPR and TPR, thus slightly variant from the mean value of the group.}
    \begin{minipage}{0.9\textwidth}
    \begin{tabularx}{\textwidth}{X<{\centering}X<{\centering}X<{\centering}X<{\centering}X<{\centering}}
        \hline
        \hline
        No.                & accuracy &    AUC &      S p-value &          B p-value \\
        \hline
        0                  &   0.9305 & 0.9660 &         0.5349 &             0.1672 \\
        1                  &   0.9365 & 0.9726 &         0.0130 &             0.7574 \\
        2                  &   0.9373 & 0.9707 &         0.4207 &             0.2046 \\
        3                  &   0.9362 & 0.9698 &         0.5278 &             0.0541 \\
        4                  &   0.9331 & 0.9671 &         0.3081 &             0.0281 \\
        5                  &   0.9362 & 0.9715 &         0.8091 &             0.0389 \\
        6                  &   0.9420 & 0.9744 &         0.3776 &             0.2883 \\
        7                  &   0.9358 & 0.9687 &         0.3732 &             0.4290 \\
        8                  &   0.9399 & 0.9734 &         0.7797 &             0.1458 \\
        9                  &   0.9352 & 0.9686 &         0.7758 &             0.5006 \\
        \hline
        mean               &   0.9363 & 0.9703 &         0.4920 &             0.2614 \\
        std                &   0.0030 & 0.0026 &         0.2372 &             0.2246 \\
        \hline
        overall            &   0.9363 & 0.9698 &              - &                  - \\
        \hline
        \hline
    \end{tabularx}
    \end{minipage}
    \label{tab:ABCNet-result}
\end{table}

\begin{figure}
    \centering
    \subfloat[10 test sets]{\includegraphics[width=0.5\textwidth]{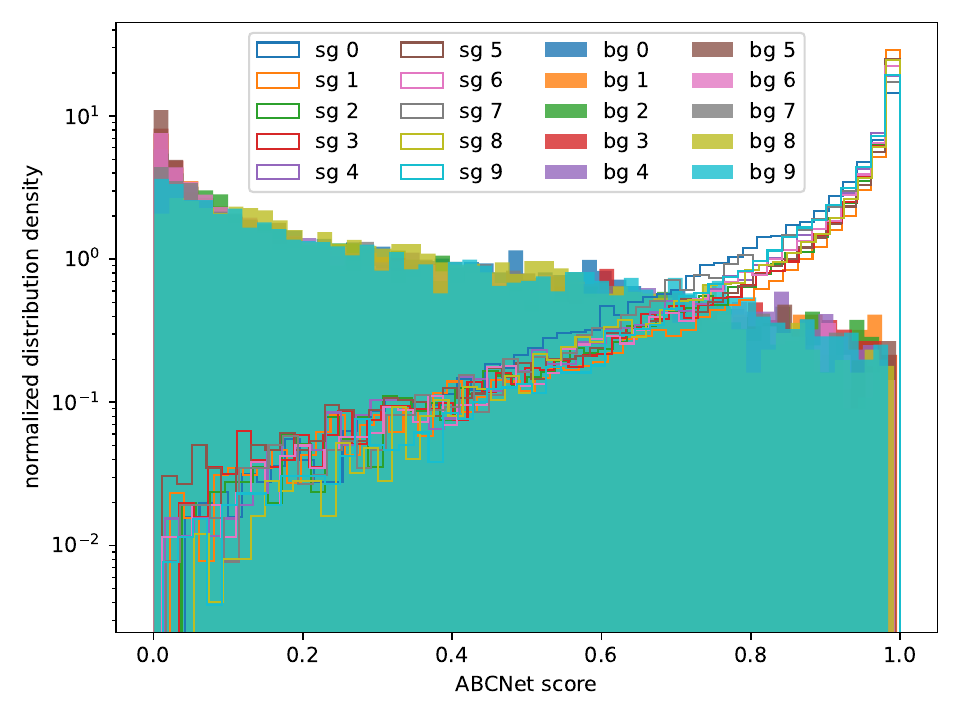}\label{fig:ABCNet-score}}
    \subfloat[training and test set 0]{\includegraphics[width=0.5\textwidth]{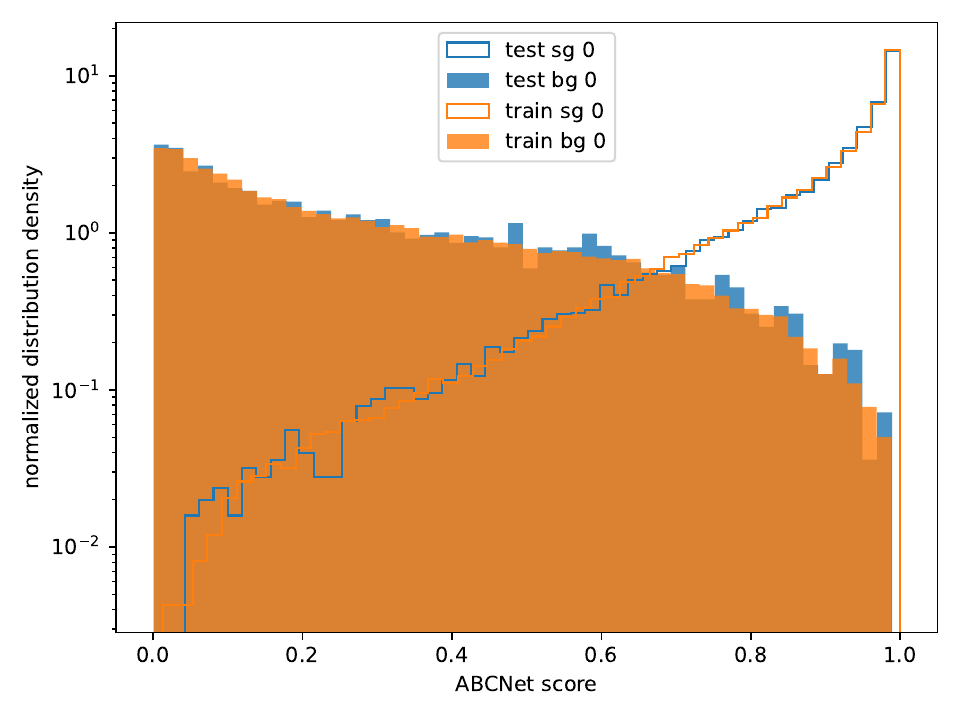}\label{fig:ABCNet-KS}}
    \caption{(a) The ABCNet score distribution histogram of the signal (sg) and background (bg) events from 10 mutually exclusive data folds. (b) The ABCNet score distributions over the training and test sets during the first round (round 0) of the 10-fold cross-validation.}
\end{figure}

Due to the insufficiency of MC samples originating from the slowness of simulation, as well as our intention to use the entire data set to set limits afterward, we equally divide the set into 10 folds for cross-validation. For each fold, the model is trained on other folds and tested on the unseen fold to give ABCNet scores. FIG. \ref{fig:ABCNet-score} visualizes the distribution of the ABCNet scores in the 10 mutually exclusive test folds. For both the signal and the background events, their distributions do not vary so much between different folds, while the signal and the background distributions do separate to a large extent. The two features show excellent generalizing and distinguishing powers of the model. To further inspect the stability and generalization of the model, TABLE \ref{tab:ABCNet-result} gives the detailed performance of the model over the 10 different folds and the overall values. For each of the 10 runs, we perform the Kolmogorov-Smirnov test to ensure that no over-fitting happens with respect to the correspondence of the ABCNet score distributions over the training and test sets for both signal and background samples, among which the result of the first run is visualized in FIG. \ref{fig:ABCNet-KS}.

\subsection{Results}
\label{sec:Results}

\begin{figure}[ht]
    \centering    
    \includegraphics[width=0.6\textwidth]{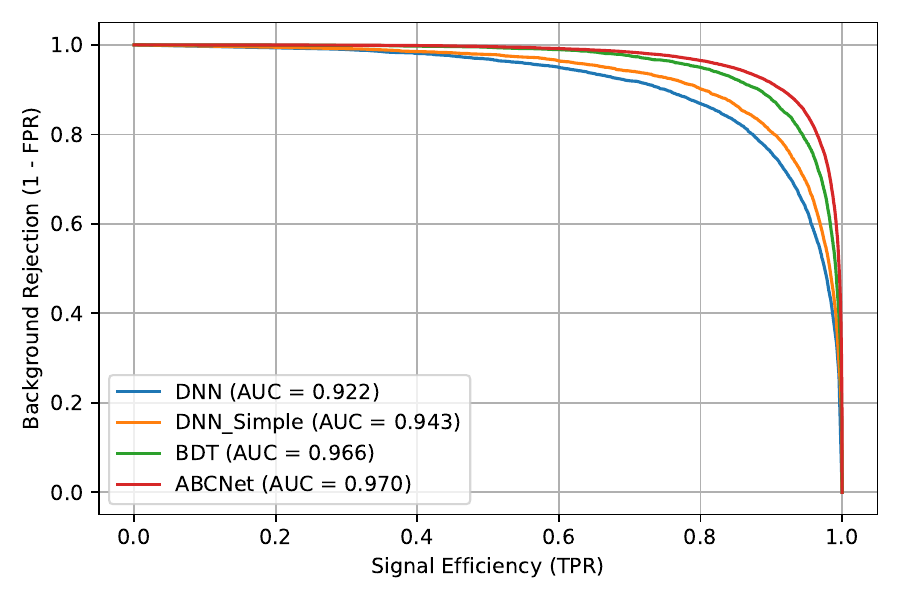}
    \caption{Comparison of background rejection versus signal efficiency for classification models: (a) BDT, (b) DNN with a single hidden layer, (c) DNN with more hidden layers, and (d) ABCNet.}
    \label{fig:ROC_DNN_BDT}
\end{figure}

Receiver Operating Characteristic (ROC) curves are shown in FIG. \ref{fig:ROC_DNN_BDT} to illustrate the performance of the classifiers. Our primary metric for comparison is the area under the ROC curve (AUC), with a higher AUC value indicating a more robust classification power across a wide range of working points. 
This metric is insightful, as it is directly connected to classification accuracy.
the quantity optimized in hyperparameter tuning, as well as other important metrics. 
The statistical significance is calculated to reveal its strong positive dependence on AUC.

Considering the AUC metric, the ABCNet has the best performance comparing to the BDT and DNNs using the same input as shown in FIG.~\ref{fig:ROC_DNN_BDT} as expected. 
%
A fully-connected DNN with many hidden layers performs even worse than the most reduced single-hidden-layer one.
%

\begin{figure}[ht]
    \includegraphics[width=0.5\textwidth,trim=.2cm .7cm 3.0cm .8cm,clip]{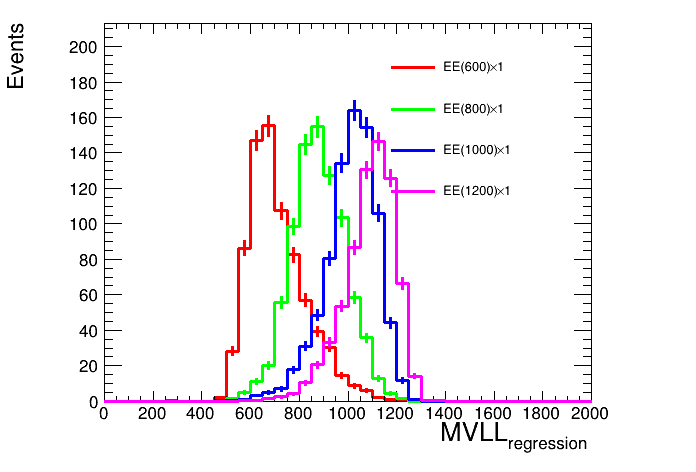}\includegraphics[width=0.5\textwidth,trim=.2cm .7cm 3.0cm .8cm,clip]{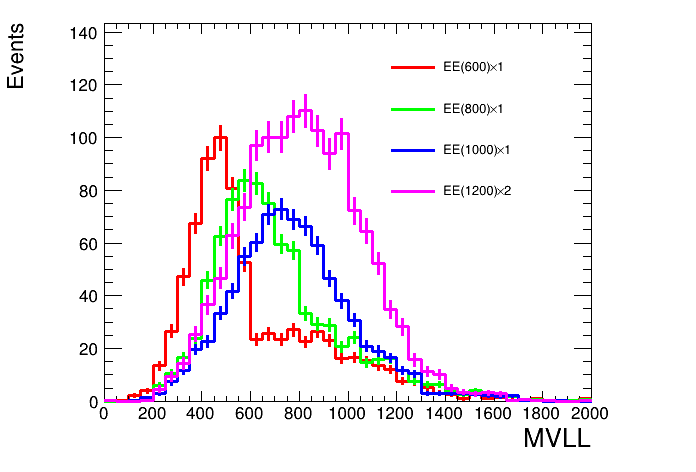}
    \caption{Comparison between the predicted \VLL mass (left) and the simplified \VLL mass reconstruction (right).}
    \label{fig:VLL_mass}
\end{figure}

We use the kinematic information of the $\tau$, jet, and \VLL candidates to predict the VLL mass, whose statistical distribution is then used to derive the observed limits on each mass point. The comparison between the predicted \VLL mass and the result reconstructed with a simplified traditional method is shown in FIG. \ref{fig:VLL_mass}, where the former fits much better.

The signal region is defined as where the discriminant (i.e., the ABCNet score) exceeds the threshold, which is chosen such that the best (i.e., strongest) 95\% CLs upper limits are obtained for all \VLL mass points. We search for peaks in regressed \VLL mass distributions by performing binned maximum likelihood fits using {\tt{HistFactory}}~\cite{Cranmer:2012sba}
and {\tt{pyhf}}~\cite{pyhf, pyhf_joss}. As observed yields are unavailable, ``Asimov'' data is used. We set the upper limit (UL) using the CLs technique~\cite{Junk:1999kv,Read:2002hq}.
Assuming the measurements performed are limited by the availability of data statistics, have well-constrained experimental systematics, and have excellent MC statistics, the background uncertainty is taken as the Poisson counting uncertainty for expected background yield in each bin.

\begin{figure}[ht]
\centering
\includegraphics[width=0.6\textwidth]{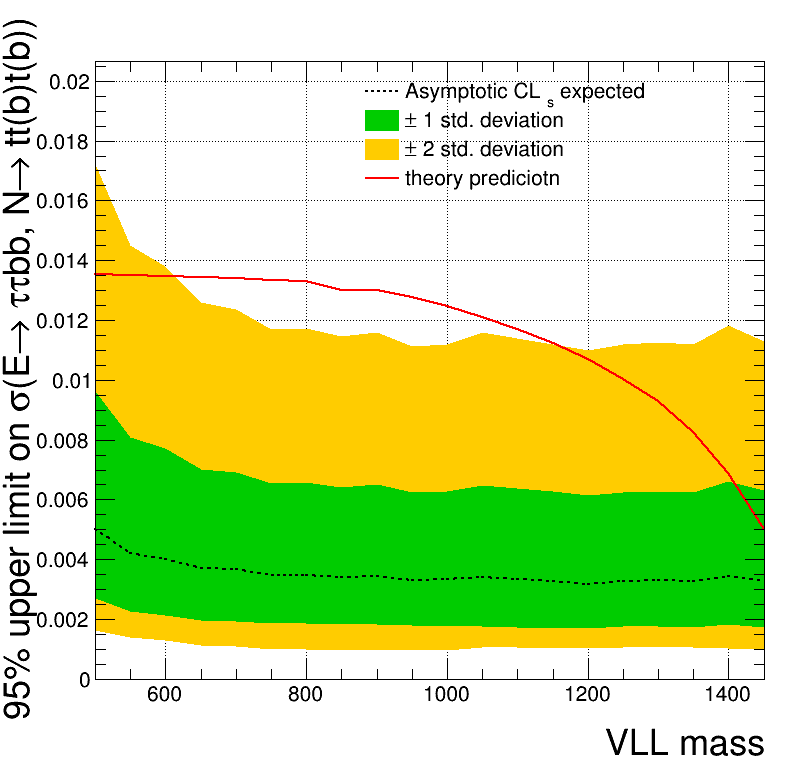}
\caption{Left: Expected and observed 95\% CL upper limits on the product of the \VLL pair production cross section from CMS experiments. Right : Expected 95\% CL upper limits on the product of the \VLL pair production cross section at a 3 TeV Muon collider.}
\label{fig:4321_cross_UL}
\end{figure}

Comparing to the results of the analysis carried out by the CMS experiment with analyzed data corresponding to an integrated luminosity of 96.5 $\mathrm{fb}^{-1}$ at a 13 TeV proton collider, results at a 3 TeV muon collider bring a significant improvement. 
In the CMS experiment, an excess of 2.8 standard deviations over the standard model background-only hypothesis was observed from the data at the representative VLL mass point of 600 GeV and no VLL masses are excluded at the 95\% confidence level.
In comparison, at a 3 TeV muon collider, \VLL masses can be excluded at the 95\% confidence level up to 1450 GeV with an integrated luminosity of only $0.01\ \mathrm{ab}^{-1}$.
The upper limit of the signal cross-section at 95\% confidence level (CL) corresponding to an integrated luminosity of $0.01\ \mathrm{ab}^{-1}$ is shown in FIG~\ref{fig:4321_cross_UL}. 

\section{Outlook and conclusions}
\label{sec:conclusions}
In this paper we investigate the potential for searching for VLL at a future high-energy muon collider. 
The 4321 model which is an ultraviolet-complete model and extends the standard model gauge groups to a larger $\text{SU}(4) \times \text{SU}(3)' \times \text{SU}(2)_L \times \text{U}(1)'$ group is considered in this work. 
In the model, a leptoquark is predicted as the primary source of lepton flavour nonuniversality, while the ultraviolet-completion predicts additional vector-like fermion families. 
In particular, VLLs are investigated via their couplings to standard model fermions through leptoquark interactions, resulting in third-generation fermion signatures. 
Final states containing at least three b-tagged jets and two $\tau$ lepton multiplicities are considered.
To improve the search sensitivity, BDT, DNN and  graph neural networks are trained to
discriminate between signal and backgrounds.
As a result, 95\% confidence level limit can be achieved for vector-like lepton mass up to 1450 GeV at a 3 TeV muon collider, with 1\% of data corresponding to the target integrated luminosity.

    
    \begin{acknowledgments}
This work is supported in part by the National Natural Science Foundation of China under Grants No. 12150005, No. 12075004 and No. 12061141002, by MOST under grant No. 2018YFA0403900. We appreciate Congqiao Li for illuminating and constructive discussions with him on neural network tuning and applications.
    \end{acknowledgments}
    
    \appendix
    \label{sec:appendix}



\begin{thebibliography}{99}
    
   \bibitem{DiLuzio:2017vat}
L.~Di Luzio, A.~Greljo and M.~Nardecchia,
Phys. Rev. D \textbf{96}, no.11, 115011 (2017)
doi:10.1103/PhysRevD.96.115011
[arXiv:1708.08450 [hep-ph]].
\bibitem{DiLuzio:2018zxy}
L.~Di Luzio, J.~Fuentes-Martin, A.~Greljo, M.~Nardecchia and S.~Renner,
JHEP \textbf{11}, 081 (2018)
doi:10.1007/JHEP11(2018)081
[arXiv:1808.00942 [hep-ph]].
 \bibitem{Greljo:2018tuh}
A.~Greljo and B.~A.~Stefanek,
Phys. Lett. B \textbf{782}, 131-138 (2018)
doi:10.1016/j.physletb.2018.05.033
[arXiv:1802.04274 [hep-ph]].

\bibitem{Blanke:2018sro}
Blanke, M. \& Crivellin, A. B Meson Anomalies in a Pati-Salam Model within the Randall-Sundrum Background. {\em Phys. Rev. Lett.}. \textbf{121}, 011801 (2018)

\bibitem{Calibbi:2017qbu}
Calibbi, L., Crivellin, A. \& Li, T. Model of vector leptoquarks in view of the B-physics anomalies. {\em Phys. Rev. D}. \textbf{98}, 115002 (2018)


\bibitem{CMS:2022cpe}
 [CMS], 
[arXiv:2208.09700 [hep-ex]].
\bibitem{Daniel20} Daniel Schulte, Nadia Pastrone, Ken Long, CERN Cour. 60 (2020) 3, 41-46.
\bibitem{Mario16} Mario Greco, Tao Han, Zhen Liu, Physics Letters B  {\bf 763} (2016) 409-415.
\bibitem{Antonio20} Antonio Costantini, et al., J. High Energ. Phys.  {\bf 2020}, 80 (2020).
\bibitem{Dario18} Dario Buttazzo, et al., J. High Energ. Phys.  {\bf 11}, 144 (2018).
\bibitem{Nazar20} Nazar Bartosik, et al., arXiv:2001.04431.
\bibitem{Sjostrand:2014zea}
T.~Sj\"ostrand, S.~Ask, J.~R.~Christiansen, R.~Corke, N.~Desai, P.~Ilten, S.~Mrenna, S.~Prestel, C.~O.~Rasmussen and P.~Z.~Skands,
Comput. Phys. Commun. \textbf{191}, 159-177 (2015)
doi:10.1016/j.cpc.2015.01.024
[arXiv:1410.3012 [hep-ph]].

\bibitem{Cacciari:2011ma}
M.~Cacciari, G.~P.~Salam and G.~Soyez,
Eur. Phys. J. C \textbf{72}, 1896 (2012)
doi:10.1140/epjc/s10052-012-1896-2
[arXiv:1111.6097 [hep-ph]].

\bibitem{Cacciari:2008gp}
M.~Cacciari, G.~P.~Salam and G.~Soyez,
JHEP \textbf{04}, 063 (2008)
doi:10.1088/1126-6708/2008/04/063
[arXiv:0802.1189 [hep-ph]].

\bibitem{deFavereau:2013fsa}
J.~de Favereau \textit{et al.} [DELPHES 3],
JHEP \textbf{02}, 057 (2014)
doi:10.1007/JHEP02(2014)057
[arXiv:1307.6346 [hep-ex]].
  
\bibitem{mucard} 
\url{https://github.com/delphes/delphes/blob/3.5.0/cards/delphes_card_MuonColliderDet.tcl}

\bibitem{muon_btag_1}
V. Di Benedetto et al., J. Inst. 13, P09004 (2018).

\bibitem{cepc_reco}
   CEPC Study Group,
   ``CEPC Conceptual Design Report: Volume 2 - Physics \& Detector,''
   arXiv:1811.10545[hep-ex]
   
\bibitem{tau_reco}
    Sirunyan, A. M. \al
    ``Performance of reconstruction and identification of $\tau$ leptons decaying to hadrons and $\nu_\tau$ in pp collisions at $\sqrt{s}=$ 13 TeV,''
    JINST 13 (2018) P10005

\bibitem{ABCNet}
Mikuni, V., \& Canelli, F. (2020). ABCNet: An attention-based method for particle tagging. \textit{The European Physical Journal Plus, 135}, 1-11.

\bibitem{GAPNet}
Chen, C., Fragonara, L. Z., \& Tsourdos, A. (2019). GAPNet: Graph attention based point neural network for exploiting local feature of point cloud. \textit{arXiv preprint arXiv:1905.08705}.

\bibitem{ParticleNet}
Qu, H., \& Gouskos, L. (2020). Jet tagging via particle clouds. \textit{Physical Review D, 101(5)}, 056019.

\bibitem{Cranmer:2012sba}
Cranmer, K., Lewis, G., Moneta, L., Shibata, A. \& Verkerke, W. HistFactory: A tool for creating statistical models for use with RooFit and RooStats.  (2012,6)

\bibitem{pyhf}
Heinrich, L., Feickert, M. \& Stark, G. pyhf: v0.7.0rc1. , https://doi.org/10.5281/zenodo.1169739, \ https://github.com/scikit-hep/pyhf/releases/tag/v0.7.0rc1

\bibitem{pyhf_joss}
Heinrich, L., Feickert, M., Stark, G. \& Cranmer, K. pyhf: pure-Python implementation of HistFactory statistical models. {\em Journal Of Open Source Software}. \textbf{6}, 2823 (2021), https://doi.org/10.21105/joss.02823

\bibitem{Junk:1999kv}
Junk, T. Confidence level computation for combining searches with small statistics. {\em Nucl. Instrum. Meth. A}. \textbf{434} pp. 435-443 (1999)

\bibitem{Read:2002hq}
Read, A. Presentation of search results: The CL(s) technique. {\em J. Phys. G}. \textbf{28} pp. 2693-2704 (2002)

\bibitem{Boronat:2016tgd}Boronat, M., Fuster, J., Garcia, I., Roloff, P., Simoniello, R. \& Vos, M. Jet reconstruction at high-energy electron–positron colliders. {\em Eur. Phys. J. C}. \textbf{78}, 144 (2018)

\bibitem{BORONAT201595}Boronat, M., Fuster, J., García, I., Ros, E. \& Vos, M. A robust jet reconstruction algorithm for high-energy lepton colliders. {\em Physics Letters B}. \textbf{750} pp. 95-99 (2015), https://www.sciencedirect.com/science/article/pii/S0370269315006565



\end{thebibliography}
    \end{document}